# The Skill-Task Matching Model: Mechanism, Model Structure, and Algorithm

*By* DA XIE AND WEIGUO YANG[*]

*We distinguished between the expected and actual profit of a firm. We proposed that, beyond maximizing profit, a firm's goal also encompasses minimizing the gap between expected and actual profit. Firms strive to enhance their capability to transform projects into reality through a process of trial and error, evident as a cyclical iterative optimization process. To characterize this iterative mechanism, we developed the Skill-Task Matching Model, extending the task approach in both multidimensional and iterative manners. We vectorized jobs and employees into task and skill vector spaces, respectively, while treating production techniques as a skill-task matching matrix and business strategy as a task value vector. In our model, the process of stabilizing production techniques and optimizing business strategies corresponds to the recalibration of parameters within the skill-task matching matrix and the task value vector. We constructed a feed-forward neural network algorithm to run this model and demonstrated how it can augment operational efficiency.*

[*] Xie: the School of labor and Human Resources, Renmin University of China, No.59 Zhongguancundajie, Haidian District, Beijing, China, 100086 (xd379349576@ruc.edu.cn). Yang (corresponding author): the School of labor and Human Resources,

Renmin University of China (weiguoyang@ruc.edu.cn). The authors declare that they have no relevant or material financial interests relating to the research described in this paper.

# 1. Introduction And Related Literature

For an extended period, the profit maximization postulate has served as a foundation for economic analyses. By evaluating the marginal returns of specific input variables, predominantly labor and capital, firms delineate optimal input-output configurations to achieve maximal profit. Cobb and Douglas (1928) introduced the Cobb-Douglas production function, modeling this mechanism based on empirical input-output data from manufacturing firms. The econometric elegance and general applicability of the Cobb-Douglas function have cemented it as a staple in Econ-101 textbooks.

However, the Cobb-Douglas production function has its limitations. It emphasizes the correlation between inputs and outputs, portraying a firm's operational process as a black box, but does not provide a comprehensive understanding of the intricate decision-making process. Moreover, it originates in the context of the Second Industrial Revolution, rendering it somewhat outdated in light of the subsequent digital technology revolution. The advent of AI algorithms, including Support Vector Machine, Convolutional Neural Networks, and Large Language Models, has significantly enhanced the computational and analytical capacities of computers, and blurs the boundary between human and machines. While economists herald these AI algorithms as formidable tools for data processing and highlight the transformative potential of AI, the Cobb-Douglas function, rooted in an earlier industrial era, seems incongruent with these algorithmic frameworks.

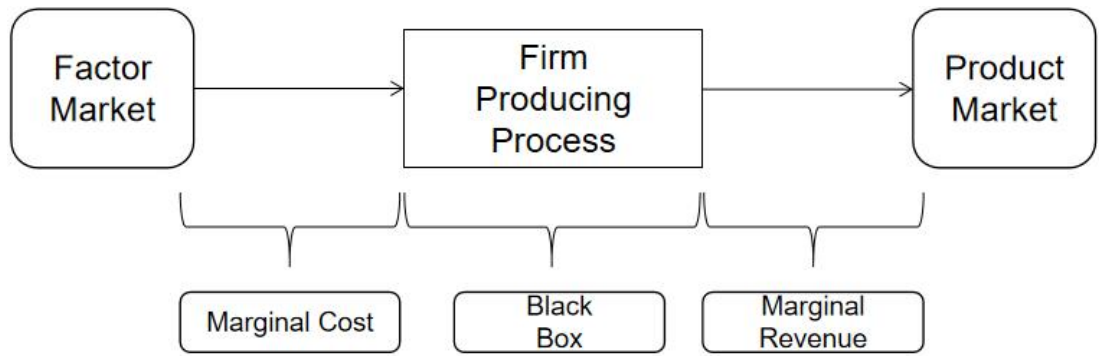

FIGURE 1: FIRM'S DECISION-MAKING BASED ON MARGINAL RETURNS

In response to these limitations, we propose a new framework to elucidate firms' decision making processes: the Iterative Mechanism. **We suggest that entrepreneurs neither engage in production haphazardly, as if partaking in a lottery game; nor do they operate the firm in a singular producing temporal interval, ceasing thereafter.** Upon discerning an unmet demand in market, entrepreneurs initially delineate an expected target. Subsequently, they disaggregate the production into distinct jobs and allocate the requisite resources to fulfill these jobs and attain the aforementioned objective. Based on actual production outcomes, entrepreneurs will adjust objectives in subsequent production cycles and modify production decisions accordingly.

**Despite an entrepreneur's capacity to ascertain the optimal factor inputs aligned with desired outputs by calculating marginal returns, the ability to flawlessly transform these optimal inputs into outputs is not innate.** Moreover, fluctuations in technology, market, and political environments necessitate a swift response in production planning. Consequently, a gap often exists between expected and realized outcomes. Entrepreneurs, therefore, continuously recalibrate production parameters through fuzzy learning to minimize this gap incrementally. This ongoing

refinement of production techniques and business strategies, honed through experiential adjustments, is defined as the Iterative Mechanism.[1]

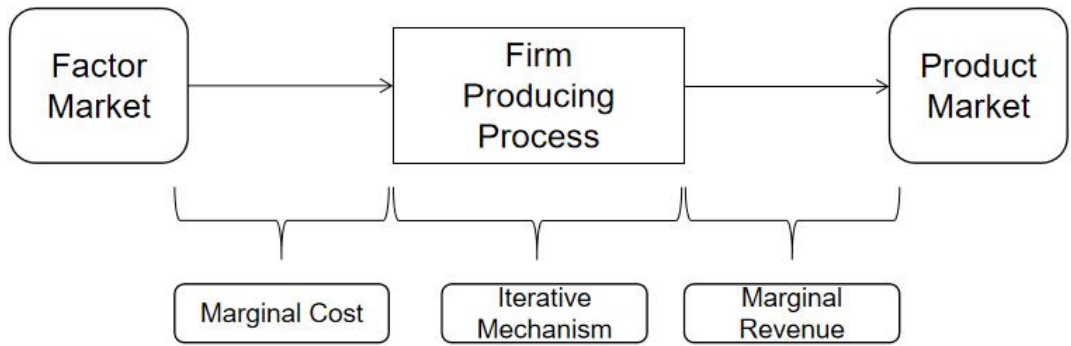

FIGURE 2: RELATIONSHIP BETWEEN THE MARGINAL RETURNS AND ITERATION MECHANISMS

Compared to Figure 1, Figure 2 introduces the Iterative Mechanism to reveal the black box of the decision process. We propose that both maximizing profit and minimizing the gap between expected and actual profit are pivotal in a firm's longitudinal aspirations. Conventionally, the profit forecasted in a given iteration corresponds to the zenith of profitability as determined by calculating the marginal cost and revenue.

Figure 3 illustrates the Iteration Mechanism:

---

[1] A quintessential example is SpaceX's Falcon 9 Heavy rocket. Elon Musk found a market gap of reusable rockets. Then he set the goal, ascertained jobs, allocated resources, engaged in iterative experimentation and refined production techniques until the actual output aligned with projections.

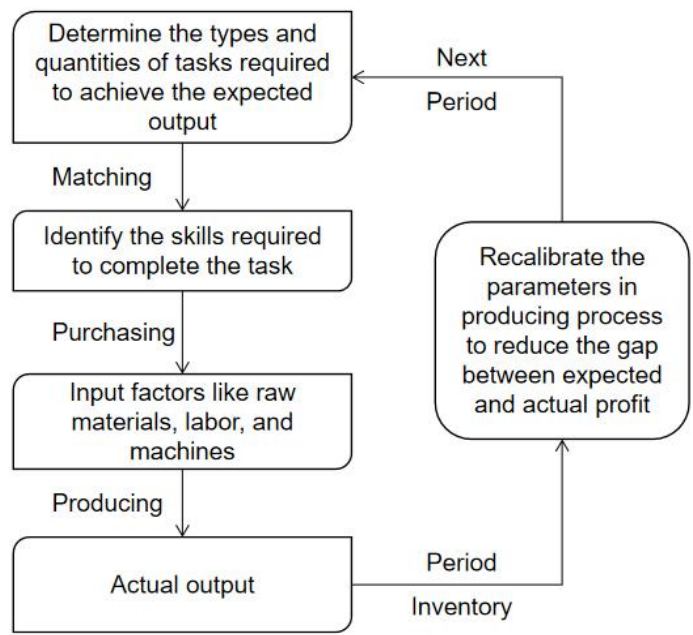

FIGURE 3: ITERATION OF THE PRODUCING PROCESS

Throughout a firm's operational lifespan, entrepreneurs continually recalibrate parameters of production to enhance the abilities to execute the blueprint. As temporal progression ensues, entrepreneurs accumulate experiential knowledge through iterative adjustments, incrementally refining and solidifying its production techniques and business strategies until it achieves operational maturity.

We develop the Skill-Task Matching Model to illustrate the iteration mechanism, by extending the task approach (Autor et al., 2003; Autor and Handel, 2013; Lise and Postel-Vinay, 2020) into a multidimensional and iterated form. Autor et al. (2003) proposed a task approach that analyzed the impact of new technology on the labor market from a task perspective. The task approach innovatively considered capital as a substitute for labor engaged in routine tasks and examined the relationships between the employment and wage levels of employees engaged in varying tasks. It marked the first model that diverged from the traditional Cobb-Douglas production function to depict the production process. Economists have extended the production function in various forms based on the task approach to explain the impact of technological change on the labor market (Autor et al., 2006; Spitz-Oener, 2006; Acemoglu and

Autor, 2011; Autor and Dorn, 2013; Autor and Handel, 2013; Goos et al., 2014; Acemoglu and Restrepo, 2017; Deming, 2017; Frey and Osborne, 2017; Lise and Postel-Vinay, 2020; Acemoglu et al., 2022).

Our model is most related to the studies of Autor and Handel (2013) and Lise and Postel-Vinay (2020). As shown in Table 1:

Table 1: The extension of task approach related to the Skill-Task Matching Model

| Production Function | Description | Author | Title | Year |
|---|---|---|---|---|
| $Y = AK^{\alpha}L^{1-\alpha}$ | Y - output; A - technology; K - capital; L - labor; $\alpha$ - output elasticity of capital and labor. | Cobb, C.W. & Douglas, P. | A Theory of Production | 1928 |
| $Y = (L_R + C)^{1-\beta}L_N^{\beta}$ $y_j = r_j^{1-\beta_j} n_j^{\beta_j}$ | Y - output; $y_j$ - output of task $j$; C - computer capital; $L_R$、$r_j$ - labor input of routine task; $L_N$、$n_j$ - labor input of non-routine task. | Autor, D.H., Levy, F. & Murnane, R.J. | The Skill Content of Recent Technological Change: An Empirical Exploration | 2003 |
| $Y_{ij} = e^{\alpha_j + \Sigma_K \lambda_{jk} \phi_{ik} + \mu_i}$ | $\alpha_j$ - threshold of position $j$; $\lambda_{jk}$ - task $k$ in position $j$; $\phi_{jk}$ - endowment that labor $i$ possesses for task $k$. | Autor, D.H. & Handel, M.J. | Putting Tasks to the Test: Human Capital, Job Tasks, and Wages | 2013 |
| $f(x,y)$ $= x_T \times [\alpha_T$ $+ \sum_{k=C,M,I}(\alpha_k y_k$ $- \kappa_k^u min\{x_k - y_k, 0\}^2$ $+ \alpha_{kk} x_k y_k)]$ | **x,y** - vectors of labor skills supply and position skills demand; T, C, M, I - generic, cognitive, manual, and interpersonal skills; $\alpha_k$ - production efficiency of task $k$; $\kappa_k^u$ - output loss when the labor's skill level and the position's skill demand do not match. | Lise, J & Postel-Vinay, F. | Multidimensional Skills, Sorting, and Human Capital Accumulation | 2020 |

The first column of Table 1 delineates the various forms of the production function. The forms in the first two rows and the last two rows of this column are distinctly different. The first two rows are predicated on the Cobb-Douglas function, while the latter two models have undergone transformation into different forms. In these scenarios, labor and capital are substituted by skills and tasks, reflecting a finer granularity. Our study follows this formation, and we prove that matching in skill and task level is more efficient than matching based on jobs and employees.

The structure of this paper is delineated as follows: The second section expounds upon the definition and structure of the Skill-Task Matching Model. The third section builds a feedforward neural network algorithm to run the model. The fourth section proves how the model improves the firm's operation efficiency. The last section provides conclusion.

## 2. The Skill-Task Matching Model

### 2.1. Definition

Definitions 1 and 2 are based on those given by Acemoglu and Autor (2011, pp.19).

Definition 1: A task is the smallest unit in the process of production activity. For any task $y_j$, $y_j \in Y$, $\boldsymbol{y_j}$ is a j-dimensional row vector, and $Y$ is the vector space of all tasks.

By definition, tasks are inseparable units during the production process. It means that each task is completely complementary, and the elasticity of substitution is $0$.[2]

---

[2] A task is the smallest process unit that transforms inputs into outputs or intermediate outputs. Therefore, if there is a substitution relationship between the two kinds of tasks, it means that these two kinds of tasks have room to be separated and are not the smallest units of production.

Definition 2: A skill is the endowment of the ability to perform tasks. For any skill $x_i$, $x_i \in X$, $\boldsymbol{x_i}$ is an i-dimensional row vector, and $X$ is the vector space of all skills.

A consensus within the economists posits that work can be conceptualized as an aggregate of tasks, and labor analogously as a composite of skills (Wong and Campion, 1991; Autor et al., 2006; Acemoglu and Autor, 2011; Marcolin et al., 2016; Ranganathan, 2023). Defination 1 and 2 move further, build vector spaces to depict work and labor.

Contrary to traditional perspectives on skill, the conception of "skill" as outlined in Definition 2 is more aligned with the task than with the human kind. In this framework, when an organization hires personnel for a particular task, "skill" becomes synonymous with human capital, with remuneration reflecting the individual's contribution in terms of human capital. In contrast, when computational systems can perform the same task and yield revenue, a corresponding return on the computer capital investment arises.[3] When skill $x_i$ is provided by the laborer, it is denoted $x_{il}$, and when it is provided by the machine, it is denoted $x_{ik}$.

Definition 3: Defining the mapping relationship $f : Y \to X$, $y_j$ as the image of $x_i$, $y_j = f(\boldsymbol{x_i})$. $f$ is the relationship between the skills required to complete a task.

In Definition 3, the mapping $f$ represents the matching relationship between the skills and tasks. By definition, $f$ is a surjection, which means that every element in set Y has a corresponding pre-image in set X.[4]

---

[3] With the development of technologies, machines can accomplish an increasing number of tasks. Knowledge-based areas such as finance, accounting, and translation, and creative tasks such as writing and painting, which were traditionally considered to be exclusively human activities, can now be completed by AI machines (Huang and Rust, 2018). As a result, the boundary between human and machine capabilities has become increasingly blurred, and machines now also possess skills.

[4] According to Definition 1, a task is the smallest unit of the production process. It must be an actual activity that occurs in the real world and have practical meaning in business production. Tasks that are currently impossible or still in laboratory conditions like "time traveling" or "human organ cloning" fall outside the purview of our research.

Define:

$$\text{matrix } A = \begin{bmatrix} a_{11} & \cdots & a_{1j} \\ \cdots & \cdots & \cdots \\ a_{i1} & \cdots & a_{ij} \end{bmatrix}, \text{ where}$$

$$\boldsymbol{x_i} \cdot A = \left[\sum_{u=1}^{i} x_u a_{u1}, \sum_{u=1}^{i} x_u a_{u2}, \ldots, \sum_{u=1}^{i} x_u a_{uj}\right] = [1,1,\ldots,1], 0 \leq a_{ij}, \ a_{ij} \in \mathbb{R}^+.$$

Matrix $A$ is a linear transformation from skill to task, represents the mapping relationship $f$. That is, the demand for different skills to complete a unit of task vector. Matrix $A$ is an invertible $i \times j$ matrix.

Definitions 1, 2, and 3 dissect the production processes. Any product's production process can be broken down into a variable number of independent tasks. These tasks aren't inherently automated; they necessitate diverse types and magnitudes of skills for completion. These requisite skills can be sourced from employees, computers, or other machinery. It's noteworthy that a single task might demand multiple skills, and conversely, a particular skill can be relevant to several tasks. The matrix $A$ delineates the degree of correlation between the skills and the tasks

Technically, the matching matrix $A$ can be seen as the transformation path of two given vectors in vector spaces $X$ and $Y$. $\boldsymbol{y_j}$ can be seen as the vectorization form of jobs and $\boldsymbol{x_i}$ is the vectorization form of employees and machines. By applying the linear transformation $A_{ij}$ to $\boldsymbol{x_i}$, vector $\boldsymbol{x_i}$ is transformed into vector $\boldsymbol{y_j}$. Conceptualizing production as a combination of skills and tasks in the form of vectors and matrices helps formalize the production process.

### 2.2. Model structure

Given these definitions, we build the Skill-Task Matching Model.

Supposing firm $\eta$ produces product $P$ in time period $t$ by the expected quantity of $Q^{tE}$. Let $\boldsymbol{y_j^{t\eta}} = \left[y_1^{t\eta}, y_2^{t\eta}, \ldots, y_j^{t\eta}\right]$ represent the task vector, which is decomposed from the work activities for producing $Q^{tE}$. When the firm considers a certain task $v$ is not essential, $y_v^{t\eta} = 0$. Matrix $A_{ij}^{t\eta} = \begin{bmatrix} a_{11}^{t\eta} & \cdots & a_{1j}^{t\eta} \\ \cdots & \cdots & \cdots \\ a_{i1}^{t\eta} & \cdots & a_{ij}^{t\eta} \end{bmatrix}$ represents the mapping relationship between skills and tasks adopted by firm $\eta$ in time period $t$. Different firms, even those producing the same product in the same quantity, have different production techniques. These differences manifest in the form of varying task quantity and structure and skill-task matching levels among firms in different time periods. Let $\boldsymbol{x_i^{t\eta}} = [x_1^{t\eta}, x_2^{t\eta}, \ldots, x_i^{t\eta}]$ represent the input skills of firm $\eta$ in period $t$. If skill $u$ is not in use, $x_u^{t\eta} = 0$. Then, the firm's actual output is $Q^t = \boldsymbol{x_i^{t\eta}} \cdot A_{ij}^{t\eta}$.[5]

Assumption 1: The expected and actual outputs are usually not equal; that is, $\boldsymbol{y_j^{t\eta}} \neq \boldsymbol{x_i^{t\eta}} \cdot A_{ij}^{t\eta}$.

Theoretically, in time period $t$, there is an ideal matching matrix $A_{ij}^{t^*\eta}$ that makes $\boldsymbol{y_j^{t\eta}} = \boldsymbol{x_i^{t\eta}} \cdot A_{ij}^{t^*\eta}$. However, in reality, a firm has no power to know exactly how well each skill and task matches. What makes it more complicated is that the task vector $\boldsymbol{y_j^{t\eta}}$ changes in each time period because of the effects of exogenous factors like weather, politics, or market fluctuations. Therefore, the firm must constantly recalibrates parameters of the skill-task matching matrix that reflect the effect of the environment on production techniques.

Hiring laborers or purchasing machines to acquire skills incurs different costs. Let $x_{ik}^{t\eta}$ represent the skills provided by machines and $x_{il}^{t\eta}$ represent the skills provided by laborers. The cost of fixed investment can be expressed as a function of the $x_{ik}^{t\eta}$, denoted as $F(x_{ik}^{t\eta})$. Let $d_i^t$ and $w_i^t$ represent the machine cost and the labor cost in period $t$. Assuming the interest rate is $r$, and the depreciation rate is $\delta$, $r$ and $\delta$ are exogenous factors. The cost function of product $P$ produced by firm $\eta$ in period $t$ is $C^{t\eta} = d_i^t x_{ik}^{t\eta} + w_i^t x_{il}^{t\eta} + \frac{F(x_{ik}^{t\eta})}{(1+r)^\delta}$.

---

[5] It is necessary to distinguish between the concepts of "variable has a value of 0" and "variable is not existing". The fact that a firm does not require a certain task or does not adopt a certain skill in the production process does not mean that the task or skill does not exist. With the changing exogenous environment, a task or a skill that currently has a value of 0 may potentially become positive in value.

Supposing the firm $\eta$ produces only one kind of product and the average price in period t is $p^t$. Let $\boldsymbol{\lambda}_j^{t\eta}$ represents the value weights assigned by the firm $\eta$ to various tasks in period $t$, $\boldsymbol{\lambda}_j^{t\eta} = [\lambda_1^{t\eta}, \lambda_2^{t\eta}, \ldots, \lambda_j^{t\eta}]$, $\lambda_j^{t\eta} \in R$.[6] Let $\hat{y}_j^{t\eta} = \sum_{k=1}^{i} x_k^{t\eta} a_{kj}^{t\eta}$ represent the actual output of task $j$. Let $I^{t\eta}$ represent expected income, $I^{t\eta} = p^t \cdot Q^{tE} = \boldsymbol{\lambda}_j^{t\eta} \cdot \boldsymbol{y}_j^{t\eta} = \sum_{u=1}^{j} y_u^{t\eta} \lambda_u^{t\eta}$. Let $\hat{I}^{t\eta}$ represent the actual income, $\hat{I}^{t\eta} = p^t \cdot Q^t = \boldsymbol{\lambda}_j^{t\eta} \cdot (\boldsymbol{x}_i^{t\eta} \cdot A_{ij}^{t\eta}) = \sum_{u=1}^{j} \hat{y}_u^{t\eta} \lambda_u^{t\eta}$.

Assumption 2: Supposing that tasks $\boldsymbol{y}_j^{t\eta}$, skills $\boldsymbol{x}_i^{t\eta}$, and total cost $C^{t\eta}$ corresponding to the expected output $Q^{tE}$ are decided at the beginning of the period $t$, whereas the actual output $Q^t = \boldsymbol{x}_i^{t\eta} \cdot A_{ij}^{t\eta}$ can only be seen at the end of period $t$.

The expected profit $\pi^{tE}$ from product $P$ by firm $\eta$ in period $t$ is: $\pi^{tE} = p^t \cdot Q^{tE} - C^{t\eta}$. The actual profit is: $\pi^t = p^t \cdot Q^t - C^{t\eta}$, and the gap between expected profit and real profit is: $\pi^{tE} - \pi^t = p^t \cdot (Q^{tE} - Q^t) = I^{t\eta} - \hat{I}^{t\eta}$. That is, given the cost generated in the beginning of the production period, the gap between the expected and actual profit is equal to the gap between the expected and actual income.

Then, we can decompose the profit gap into distinct tasks:

$$\boldsymbol{R}_j^{t\eta} = [r_1^{t\eta}, r_2^{t\eta}, \ldots, r_j^{t\eta}] = I^{t\eta} - \hat{I}^{t\eta} = \boldsymbol{\lambda}_j^{t\eta}(\boldsymbol{y}_j^{t\eta} - \boldsymbol{x}_i^{t\eta} \cdot A_{ij}^{t\eta}) \tag{1}$$

---

[6] Different firms have different business strategies such as operating policies, advertising styles and company cultures. These differences were depicted as different $\lambda_j^{t\eta}$.

Equation (1) illustrates the gap between expected and actual profits at the task level, where $r_j^{t\eta} = \lambda_j^{t\eta}(y_j^{t\eta} - \sum_{u=1}^{i} x_u^{t\eta} a_{uj}^{t\eta})$ represents the gap between expected profit and real profit brought by task $j$ in period $t$. If $r_j^{t\eta} > 0$ indicates that the real profits are lower than expected, and if $r_j^{t\eta} < 0$ means that the real profits are higher than expected. In either case, the firm would recalibrate the parameters, and continues its production process in next period. Equation (1) implies that the gap between the expected and actual profit is equal to the difference between the expected and actual income under the Assumption 2.

Assumption 3: Firms try to fine-tune production parameters by fuzzy learning.

Upon observing a gap between expected and actual profits, firms would establish a loss function and recalibrate the production parameters to reduce the value of this loss function. This process is executed by entrepreneurs or engineers through fuzzy learning. Define the loss function[7]:

$$E_A^{t\eta} = \frac{1}{2}\sum_{s=1}^{j}(\hat{y}_s^{t\eta} - y_s^{t\eta})^2;$$

$$E_\lambda^{t\eta} = \frac{1}{2}(\hat{I}^{t\eta} - I^{t\eta})^2.$$

Assuming the learning rate are $\theta_A$ and $\theta_\lambda$, $\theta_A \in (0,1)$, $\theta_\lambda \in (0,1)$.[8] Then, the updated estimation formula for $a_{ij}^{t\eta}$ is $a_{ij}^{t+1,\eta} = a_{ij}^{t\eta} + \Delta a_{ij}^{t\eta}$, where:

$$\Delta a_{ij}^{t\eta} = \theta_A \cdot \frac{\partial E_A^{t\eta}}{\partial a_{ij}^{t\eta}} = \theta_A \cdot \frac{\partial E_A^{t\eta}}{\partial \hat{y}_j^{t\eta}} \cdot \frac{\partial \hat{y}_j^{t\eta}}{\partial a_{ij}^{t\eta}} = \theta_A \cdot (\hat{y}_j^{t\eta} - y_j^{t\eta}) \cdot x_i^{t\eta} \quad (2)$$

The updated estimation formula for $\lambda_j^{t\eta}$ is $\lambda_j^{t+1,\eta} = \lambda_j^{t\eta} + \Delta \lambda_j^{t\eta}$, where:

$$\Delta \lambda_j^{t\eta} = \theta_\lambda \cdot \frac{\partial E_\lambda^{t\eta}}{\partial \lambda_j^{t\eta}} = \theta_\lambda \cdot \frac{\partial [\sum_{u=1}^{j} \lambda_u^{t\eta}(\hat{y}_u^{t\eta} - y_u^{t\eta})]}{\partial \lambda_j^{t\eta}} = \theta_\lambda \cdot (\hat{y}_j^{t\eta} - y_j^{t\eta}) \quad (3)$$

---

[7] We define the loss function as an ordinary second order moment. Other types of loss functions could also be specified.

[8] The learning rate represents how "fuzzy" the recalibrating process is; the larger it is, the fuzzier the process.

Equation (2) and (3) demonstrate how to fine-tune production parameters of each task. Identifying tasks where there is a gap between expected and actual outputs, as well as setting an appropriate learning rate, are heavily relied on the entrepreneur's mastery level of the parameters within the matching matrix and task value vector.

Statement 1: If $R_j^{t\eta} \neq 0$, then $a_{ij}^{t+1,\eta} = a_{ij}^{t\eta} + \Delta a_{ij}^{t\eta}$, $\lambda_j^{t+1,\eta} = \lambda_j^{t\eta} + \Delta \lambda_j^{t\eta}$, firms would assign $A_{ij}^{t+1,\eta}$ to $A_{ij}^{t,\eta}$, assign $\lambda_j^{t+1,\eta}$ to $\lambda_j^{t,\eta}$, $t = t + 1$, where $A_{ij}^{t+1,\eta} = A_{ij}^{t\eta} + \Delta A_{ij}^{t\eta} = (a_{i1}^{t+1,\eta}, a_{i2}^{t+1,\eta}, \ldots, a_{ij}^{t+1,\eta})$.

Statement 1 is the core of iteration mechanism. It could be shown as a flow chart in Figure 4:

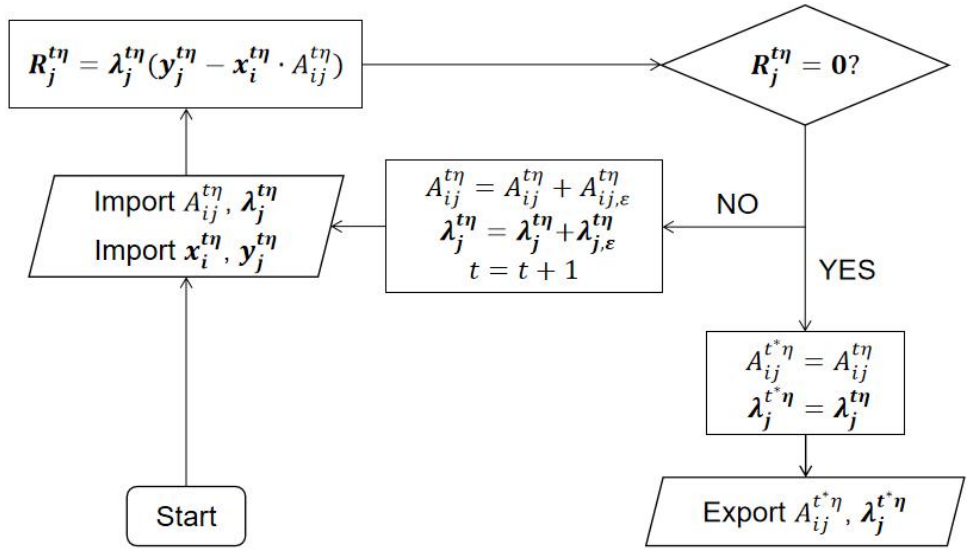

FIGURE 4: FLOW CHART OF ITERATION

Entrepreneurs will set initial parameter values of the matching matrix and value vector, and incorporate anticipated output-related task vectors and the skill vectors as inputs when $t = 0$. As long as there is a gap between the expected and actual profit, entrepreneurs will locate the task vector by the loss function, and recalibrate the parameters by fuzzy learning. In the subsequent period, entrepreneurs continue to narrow the gap between expected and actual outcomes. The iteration process will keep going on until the firm masters the production techniques and business strategies associated with that task. At this juncture, the production parameters for that task are rendered as fixed.

### 3. Constructing AI algorithm for the recalibrating process

Modern algorithms offer insights into the cognitive processes of the human brain, enable us to represent the entrepreneur's continuous decision-making process in algorithmic form. We build a feedforward neural network algorithm to simulate the Statement 1. The recalibrating of production parameters can be divided into two steps: first, recalibrating $A_{ij}^{t\eta}$ to $A_{ij}^{t^*\eta}$ to the point where the number of skills input converges to the number of tasks projected; second, recalibrating $\lambda_j^{t\eta}$ to $\lambda_j^{t^*\eta}$ to the point where the value of task output is close to expected income. Figure 5 illustrates the first step of this feedforward neural network:

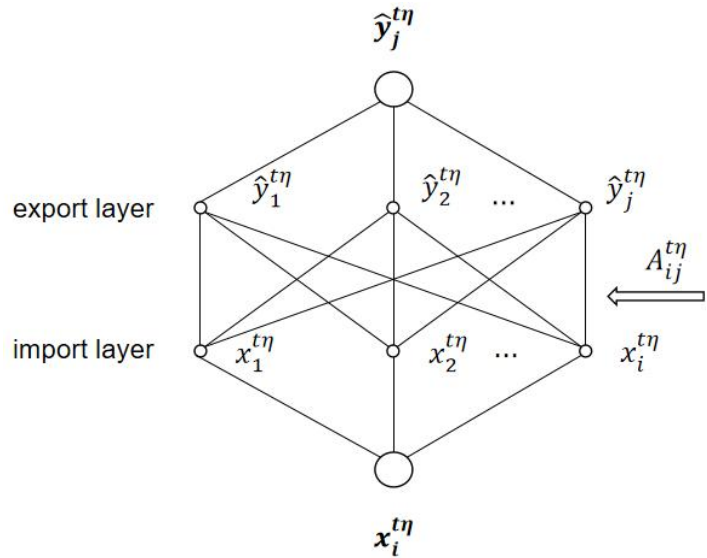

FIGURE 5: USING FEEDFORWARD NEURAL NETWORK TO OBTAIN $A_{ij}^{t^*\eta}$

Assuming the bias term is zero and the activation function is 1. The training set is $S_A = \{(x_i^{1\eta}, y_j^{1\eta}), (x_i^{2\eta}, y_j^{2\eta}), \ldots, (x_i^{t\eta}, y_j^{t\eta})\}$. For the t-th training sample, the inputs are $x_i^{t\eta}$ and $y_j^{t\eta}$. The neural network output is $\hat{y}_j^{t\eta} = (\hat{y}_1^{t\eta}, \hat{y}_2^{t\eta}, \ldots, \hat{y}_j^{t\eta})$, where $\hat{y}_j^{t\eta} = \sum_{k=1}^{i} x_k^{t\eta} a_{kj}^{t\eta}$.

The content of the algorithm is:

Import: training set $S_A = \{(x_i^{v\eta}, y_j^{v\eta})\}_{v=1}^{t}$;

Given learning rate $\theta_A$.

Process:

1. Initializing the numbers in $A_{ij}^{v\eta}$

2. Repeat:

3. For all $(x_i^{v\eta}, y_j^{v\eta}) \in S_A$, do

4. Calculating the output $\hat{y}_j^{v\eta}$ of the current sample

5. Refreshing the weighting matrix $A_{ij}^{v\eta}$ by equation (2)

6. If $\Delta a_{ij}^{t\eta} = 0$, End for

7. Export results

The second step is illustrated in Figure 6:

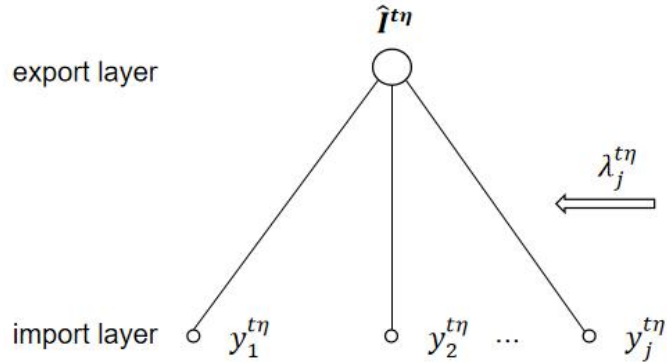

FIGURE 6: USING FEEDFORWARD NEURAL NETWORK TO OBTAIN $\lambda_j^{t^*\eta}$

Algorithm for acquiring $\lambda_j^{t^*\eta}$ is easier than for $A_{ij}^{t^*\eta}$. The assumptions and import sets remain unchanged. Let $I^{t\eta}$ represents the expected output for the t-th period for the t-th training sample $(y_j^{t\eta}, I^{t\eta})$, the input is $y_j^{t\eta}$ and $I^{t\eta}$. The neural network output is $\hat{I}^{t\eta} = \sum_{u=1}^{j} y_u^{t\eta} \lambda_u^{t\eta}$.

The content of algorithms is:

Import: training set $S_\lambda = \{y_j^{t\eta}, I^{t\eta})\}_{v=1}^{t}$;

Given learning rate $\theta_\lambda$.

Process:

1. Initializing the numbers in $\lambda_j^{t\eta}$

2. Repeat:

3. For all $(y_j^{t\eta}, I^{t\eta}) \in S_\lambda$, do

4. Calculating the output $\hat{I}^{t\eta}$ of the current sample

5. Refreshing the weighting vector $\lambda_j^{t\eta}$ by equation (3)

6. If $\Delta \lambda_j^{t\eta} = 0$, End for

7. Export results

These two steps show how to apply neural network algorithms on acquiring optimal production parameters. The outputs, $A_{ij}^{t^*\eta}$ and $\lambda_j^{t^*\eta}$, are the production parameters in a mature stage. By accumulating sufficient data, AI algorithm could simulate the human decision making process, digitalize the process of recalibrating production parameters from physical to digital space, and significantly compress the time cost to enhance execution capabilities.

### 4. How the model improves the operation efficiency

Historically, technological limitations impeded the granular decomposition of work into distinct tasks (Lazear, 1995, pp.87). Given the situation, the recalibration of parameters to realize anticipated profits was predominantly contingent upon the matching level of occupations and employees. In the wake of the AI revolution, emergent recognition and computational tools have empowered firms to dissect work at an intricate task level and improve the operating efficiency.

### 4.1 Improving matching efficiency

Proposition 1. Applying the Skii-Task Matching Model could enhance the firm's matching efficiency.

Proof: Assuming the composition of tasks of the occupation Q is $y_Q^{t\eta} = [y_1^{t\eta}, y_2^{t\eta}, \ldots, y_q^{t\eta}]$, the structure of the employee L's skill is $L_l^t = [x_1^{tl}, x_2^{tl}, \ldots, x_i^{tl}]$, and the skill - task matching matrix established by the firm is $A_{iQ}^{t\eta}$:

$$A_{iQ}^{t\eta} = \begin{bmatrix} a_{11}^{t\eta} & \cdots & a_{1q}^{t\eta} \\ \cdots & \cdots & \cdots \\ a_{i1}^{t\eta} & \cdots & a_{iq}^{t\eta} \end{bmatrix}$$

Assuming the task value weight vector is $\lambda_Q^{t\eta} = [\lambda_1^{t\eta}, \lambda_2^{t\eta}, \ldots, \lambda_q^{t\eta}]$. The skill levels and the workforce structure are regarded as constant during one time interval.[9] Subsequently, the gap between the expected and actual profit is:

$$R_Q^{lQ} = [r_1^{lQ}, r_2^{lQ}, \ldots, r_q^{lQ}] = \lambda_Q^{t\eta}(y_Q^{t\eta} - L_l^t \cdot A_{iQ}^{t\eta})$$

$$= [\lambda_1^{t\eta}(y_1^{t\eta} - \sum_{u=1}^{i} x_u^{tl} a_{u1}^{t\eta}), \lambda_2^{t\eta}(y_2^{t\eta} - \sum_{u=1}^{i} x_u^{tl} a_{u2}^{t\eta}), \ldots, \lambda_q^{t\eta}(y_q^{t\eta} - \sum_{u=1}^{i} x_u^{tl} a_{uq}^{t\eta})].$$

Assuming that $r_Q^{lQ} = max[0, \lambda_q^{t\eta}(y_q^{t\eta} - \sum_{u=1}^{i} x_u^{tl} a_{uq}^{t\eta})]$, which means when workers' skill levels surpass the requisites of their tasks, the task output cannot be further enhanced.

Case 1. All the employees' skills meet every task's requirement

In this case, $R_Q^{lQ} \geq 0$, implis that the employee could execute all the tasks without any decrease in output, leading to the realization of the maximum output, no gap between the expected and actual profit. There's no difference between matching by occupations and tasks as well. However, it underutilizes the competencies of laborers in intricate undertakings, leading to overeducation and the squandering of human resources, damages the profit.

---

[9] In other words, the accumulation of human capital acquired by employees through learning on the job is treated as a discrete function of time, with t as the smallest time interval.

Case 2. The employee's skills meet only part of tasks' requirements

In most cases, employees are specialized in specific skills pertinent to particular occupational functions. It means that there are misalignments between task structures and the employees' skills (Autor and Handel, 2013). If employee L's skills do not meet the prerequisites of task $u$, $r_u^{lQ} < 0$, the actual output of the occupation is $\sum_{v=1}^{q}(\sum_{u=1}^{i} x_u^{tl} a_{uv}^{t\eta} \cdot \lambda_v^{t\eta})$.[10] Assuming L's payments correlate directly with the tangible output. Before the AI revolution, the allocation of workforce into production is based on matching between occupation and employee. Therefore, the objective function of firm $\eta$ is:

$$\max \sum_{v=1}^{q}(\sum_{u=1}^{i} x_u^{tl} a_{uv}^{t\eta} \cdot \lambda_v^{t\eta})$$

Nowadays, firms are capable of matching skills of employees and tasks of pccupations in smaller granularity, the objective function firm $\eta$ is:

$$q \cdot [max(x_u^{tl} a_{uv}^{t\eta} \cdot \lambda_v^{t\eta})]$$

Obviously:

$$q \cdot [max(x_u^{tl} a_{uv}^{t\eta} \cdot \lambda_v^{t\eta})] \geq \max \sum_{v=1}^{q}(\sum_{u=1}^{i} x_u^{tl} a_{uv}^{t\eta} \cdot \lambda_v^{t\eta}) \qquad (4)$$

**Given a set of variables, the summation of their individual maxima is always greater than or equal to the maxima of their total sum.** As evidenced by equation (4), by applying the Skill-Task Matching Model, a firm could enhance the matching efficiency of work and labor.

---

[10] $\sum_{u=1}^{i} x_u^{tl} a_{uv}^{t\eta}$ denotes the actual output of task v, $\lambda_v^{t\eta}$ denotes the value of task v, $\sum_{v=1}^{q}(\sum_{u=1}^{i} x_u^{tl} a_{uv}^{t\eta} \cdot \lambda_v^{t\eta})$ denotes the actual output value of all tasks in occupation Q.

## 4.2 Improving iterating efficiency

From a production standpoint, shortening the iteration cycle enables a firm to adapt to environmental fluctuation more swiftly. The duration of an iteration cycle is contingent on the time required to complete occupations and the firm's parallel management capabilities.

Proposition 2. Applying the Skill-Task Matching Model could enhance the iteration efficiency.

Proof: Assuming that the number of the position Q in the enterprise in period $t$ is $D_Q^{t\eta}$, the task vector of Q is $y_Q^{t\eta} = \left[y_{Q1}^{t\eta}, y_{Q2}^{t\eta}, \ldots, y_{Qq}^{t\eta}\right]$, asumming the time required to complete $y_{Qq}^{t\eta}$ is $w_{Qq}^{t\eta}$, $w_{Qq}^{t\eta} \in \mathbb{R}^+$, then the time required to complete the position Q is given by:

$$w_Q^{t\eta} \in (max(w_{Q1}^{t\eta} \cdot y_{Q1}^{t\eta}, w_{Q2}^{t\eta} \cdot y_{Q2}^{t\eta}, \ldots, w_{Qq}^{t\eta} \cdot y_{Qq}^{t\eta}), \sum_{m=1}^{q} w_{Qm}^{t\eta} \cdot y_{Qm}^{t\eta}) \qquad (5)$$

Equation (5) denotes that the time spent on an occupation falls between the time taken by the longest task and it that taken by the sum of all tasks. Supposing the firm sets up $n$ occupations, then the iteration cycle $t$ of the enterprise $n$ to complete the production of the product G is:

$$t \in (max(w_1^{t\eta} \cdot D_1^{t\eta}, w_2^{t\eta} \cdot D_2^{t\eta}, \ldots, w_n^{t\eta} \cdot D_n^{t\eta}), \sum_{k=1}^{n} w_k^{t\eta} \cdot D_k^{t\eta}).$$

That is, the duration of a production iteration stage is between the time taken by the longest occupation and it that taken by the sum of all occupations.

Supposing the firm has $J$ tasks in total, the type and total number of tasks remain unchanged. It means that $J = \sum_{s=1}^{n} \sum_{z=1}^{q} y_{sz}^{t\eta} \cdot D_s^{t\eta}$. The iteration period of the production product G is:

$$t \in (max(w_1^{t\eta} \cdot y_1^{t\eta}, w_2^{t\eta} \cdot y_2^{t\eta}, \ldots, w_J^{t\eta} \cdot y_J^{t\eta}), \sum_{v=1}^{J} w_v^{t\eta} \cdot y_v^{t\eta}).$$

Since the type and total number of tasks remain unchanged, by combining equation (5), it could be obtained:

$$\sum_{k=1}^{n} w_k^{t\eta} \cdot D_k^{t\eta} \geq \sum_{u=1}^{k} \sum_{m=1}^{q} w_{um}^{t\eta} \cdot y_{um}^{t\eta} = \sum_{v=1}^{J} w_v^{t\eta} \cdot y_v^{t\eta} \qquad (6)$$

$$max(w_1^{t\eta} \cdot D_1^{t\eta}, w_2^{t\eta} \cdot D_2^{t\eta}, \ldots, w_n^{t\eta} \cdot D_n^{t\eta}) \geq max(w_1^{t\eta} \cdot y_1^{t\eta}, w_2^{t\eta} \cdot y_2^{t\eta}, \ldots, w_J^{t\eta} \cdot y_J^{t\eta}) \qquad (7)$$

Equations (6) and (7) indicate that **the iteration period at the skill-task level is always shorter than or equal to that at the employee-occupation level**. Firms with shorter iteration cycles could adapt to environmental changes more effectively and adjust their strategies with greater flexibility.

## 5. Conclusion

We proposed the iteration mechanism to illustrate how a firm bridges the gap between expected and actual profit. We built the Skill-Task Matching Model, which extends the task approach both iteratively and multi-dimensionally, to depict this mechanism. We vectorized work into a task vector space and labor (machine) into a skill vector space. The matching matrix between skills and tasks quantificats the firm's production techniques, while the task value vector quantificats the firm's business strategies. By accumulating fuzzy learning results, a firm's production techniques and strategies are facilitated by the process of recalibrating the parameters of matching matrix and task value vector iteratively. Therefore, the production process evolves from approximations to precision, and actural profits incrementally align with its expected profits. We developed a feedforward neural network algorithm to run the model, and prove that its application could enhance the operation efficiency of the firm.

The model we proposed revealed the black box of a firm's decision making process, and introduced AI algorithms into the economic models. By recognizing more tasks and skills and accumulating more data, further research could be conducted to analyze the relationship between certain tasks and skills, returns to skills or skills bundles, the application of AI algorithms, and other topics in micro-economics.